\documentclass[conference]{IEEEtran}
\IEEEoverridecommandlockouts
\usepackage{amsmath,amssymb,amsfonts}
\usepackage{algorithmic}
\usepackage{graphicx}
\usepackage{textcomp}
\usepackage{xcolor}

\usepackage[caption=false]{subfig}
\usepackage[hidelinks]{hyperref}
\usepackage{mathtools}
\usepackage{balance}

\usepackage[backend=biber,style=ieee,natbib=true]{biblatex} 
\usepackage{braket}
\usepackage[textsize=tiny]{todonotes}

\newcommand{\B}[1]{\boldsymbol{#1}}
\newcommand\norm[1]{\left\Vert#1\right\Vert}

\begin{document}

\title{Challenges for Reinforcement Learning in\\ Quantum Circuit Design\thanks{This is an extended version of \citep{Altmann24-QCD}, accepted for publication at the 2024 IEEE International Conference on Quantum Computing and Engineering (QCE). }}

\author{
\IEEEauthorblockN{Philipp Altmann}
\IEEEauthorblockA{\textit{LMU Munich}\\\quad Munich, Germany\quad\\
philipp.altmann@ifi.lmu.de}
\and
\IEEEauthorblockN{Jonas Stein}
\IEEEauthorblockA{\textit{LMU Munich}\\\quad Munich, Germany\quad}
\and
\IEEEauthorblockN{Michael Kölle}
\IEEEauthorblockA{\textit{LMU Munich}\\\quad Munich, Germany\quad}
\and
\IEEEauthorblockN{Adelina Bärligea}
\IEEEauthorblockA{\textit{TU Munich}\\\quad Munich, Germany\quad}
\and
\IEEEauthorblockN{Maximilian Zorn}
\IEEEauthorblockA{\textit{LMU Munich}\\\quad Munich, Germany\quad}
\and
\IEEEauthorblockN{Thomas Gabor}
\IEEEauthorblockA{\textit{LMU Munich}\\\quad Munich, Germany\quad}
\and
\IEEEauthorblockN{Thomy Phan}
\IEEEauthorblockA{\textit{University of Southern California}\\Los Angeles, USA}
\and
\IEEEauthorblockN{Sebastian Feld}
\IEEEauthorblockA{\textit{Delft University of Technology}\\Delft, The Netherlands}
\and
\IEEEauthorblockN{Claudia Linnhoff-Popien}
\IEEEauthorblockA{\textit{LMU Munich}\\\quad Munich, Germany\quad}
}

\maketitle

\begin{abstract}
Quantum computing (QC) in the current NISQ era is still limited in size and precision. Hybrid applications mitigating those shortcomings are prevalent to gain early insight and advantages. Hybrid quantum machine learning (QML) comprises both the application of QC to improve machine learning (ML) and ML to improve QC architectures. This work considers the latter, leveraging reinforcement learning (RL) to improve quantum circuit design (QCD), which we formalize by a set of generic objectives. Furthermore, we propose \texttt{qcd-gym}, a concrete framework formalized as a Markov decision process, to enable learning policies capable of controlling a universal set of continuously parameterized quantum gates. Finally, we provide benchmark comparisons to assess the shortcomings and strengths of current state-of-the-art RL algorithms. 
\end{abstract}


\section{Introduction}

Quantum circuits comprise a combination of unitary compositions, or quantum gates, executed on a quantum computer to alter a particular quantum state.
Mostly driven by theoretical promises of achieving exponential speed-ups in computational tasks such as integer factorization \cite{Shor.1999} and solving linear equation systems \cite{harrow2009quantum}, quantum computing (QC) has become a rapidly growing field of research. 
However, the current era of \textit{noisy intermediate-scale quantum} (NISQ) hardware \citep{Preskill.2018} poses considerable limitations that prevent the realization of these speed-ups.
Moreover, prevalent circuit architectures are mostly handcrafted.
Therefore, the central task of \textit{quantum circuit design} (QCD) consists of finding a sequence of quantum gates to achieve a specific objective, potentially given (hardware) limitations.
We mainly consider two categories, architecture search and circuit optimization, more specifically, the preparation of arbitrary states and the composition of unitary operations.
\begin{figure}\centering
\includegraphics[width=0.48\textwidth]{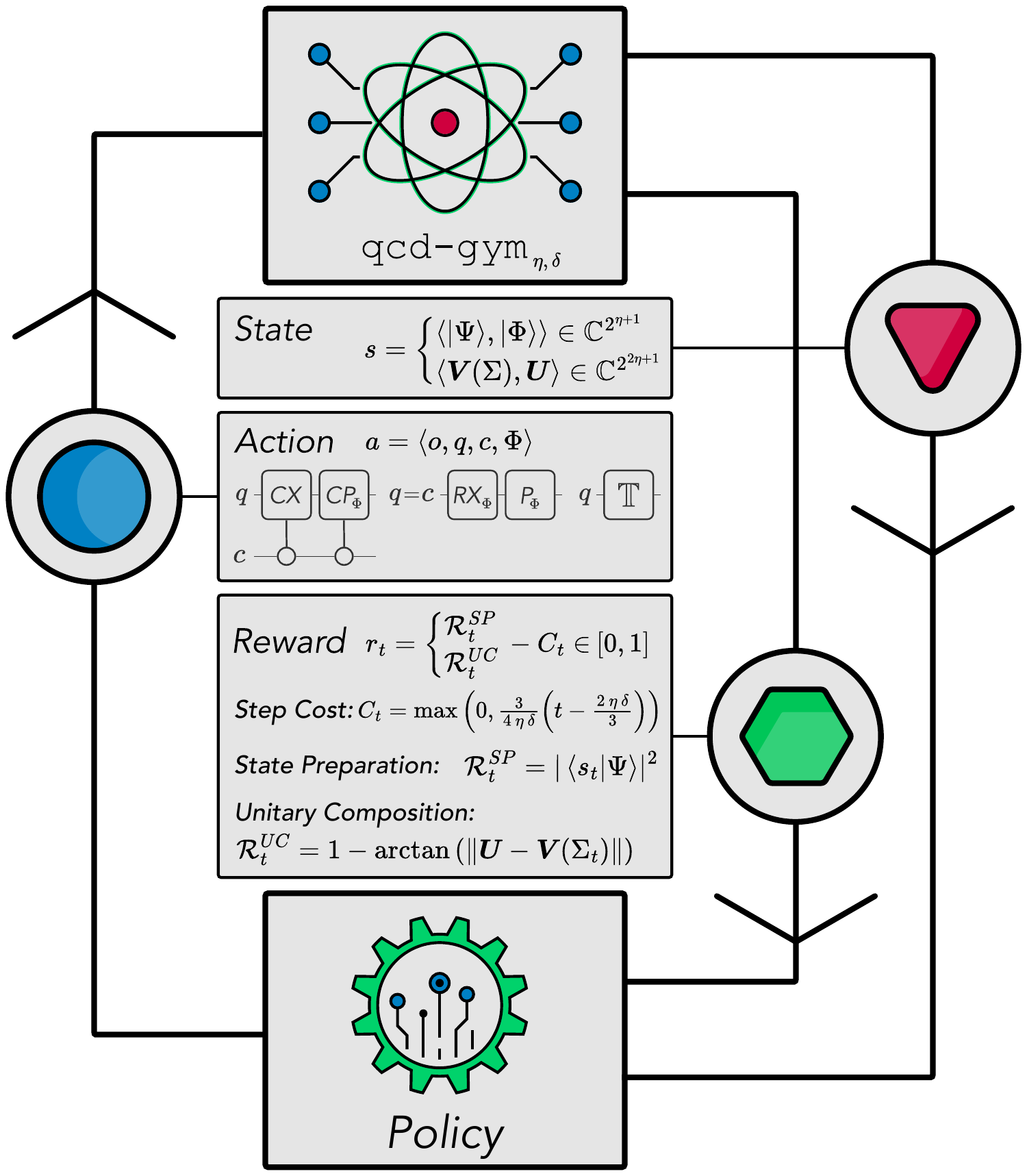}
\caption{\texttt{qcd-gym} for $\eta$ qubits with depth $\delta$ for generating a sequence $\Sigma$ of continuously parameterized operation $a$ (blue) to optimally resemble a target state $\ket{\Phi}$ or unitary $\boldsymbol{U}$ in a single optimization loop, based on the observed state $s$ (red) and the reward $r_t$ (green). }\label{fig:circuit-designer}\vspace{-1em}
\end{figure}
Among other ML approaches, we find \textit{reinforcement learning} (RL) to be specifically suited for these objectives, learning tasks in various discrete and continuous sequential decision-making problems with temporal dependence, without the need for granular specification.
Also, sophisticated mechanisms balancing exploration and exploitation provide a helpful foundation for discovering viable architectures.
Through this research, we aim to provide comprehensive insight into various objectives associated with quantum circuit design, fostering collaboration and innovation within the AI research community.
However, in contrast to most related work, we approach this topic bottom-up by dividing problems of interest into their core objectives, aiming to shed light on the challenges that current RL algorithms face in this field of research and laying the groundwork for all-embracing automated circuit design.
Overall, we provide the following contributions:
\begin{itemize}
\item We formulate core objectives for reinforcement learning arising from quantum circuit design (QCD).
\item We define \texttt{qcd-gym}, a generic RL framework for QCD to learn executing parameterized operations (cf. Fig.~\ref{fig:circuit-designer}).
\item We demonstrate the challenges current state-of-the-art RL approaches face in QCD.
\end{itemize}

\section{Background}

\subsection{Quantum Computing}
First, we provide a brief overview of the basic concepts of quantum computing. 
For more detailed information, see \citet{nielsen2010quantum}. 
A quantum computer is built from qubits (wires) to execute a quantum circuit defined by a sequence of elementary operations (quantum gates) to carry and manipulate the quantum state. 
Unlike classical bits, which can be in the state $0$ or $1$, qubits can take on any superposition of these two, so their state can be represented as: 
\begin{equation}
\alpha \ket{\B{0}} + \beta \ket{\B{1}} = \alpha \begin{bmatrix} 1 \\ 0 \end{bmatrix} + \beta \begin{bmatrix} 0 \\ 1 \end{bmatrix} = \begin{bmatrix} \alpha \\ \beta \end{bmatrix} \ \in \mathbb{C}^2.
\label{eq:QubitState}
\end{equation}
$\ket{\B{0}}$ and $\ket{\B{1}}$ are quantum computational basis states of the two-dimensional Hilbert space spanned by one qubit, and $|\alpha|^2 + |\beta|^2=1$. A two-qubit state can be represented by $\ket{\B{a} \B{b}} = \ket{\B{a}} \ket{\B{b}} = \ket{\B{a}} \otimes \ket{\B{b}} \in \mathbb{C}^{2^2}$, with the $2^2$ computational basis states $\ket{\B{0}\B{0}}$, $\ket{\B{0}\B{1}}$, $\ket{\B{1}\B{0}}$, and $\ket{\B{1}\B{1}}$. 
Following this formulation, quantum gates can be understood as unitary matrix operations $\B{U}\in\mathbb{C}^{2^n\times2^n}$ applied to the state vector of the $n$ qubits, on which the gate acts. 
Their only condition of being unitary (i.e., $\B{U}\B{U^\dag} = \B{I}$) ensures that the operations inherent in them are reversible. 
With the identity $\B{I}$, the most elementary single-gate operations contain: 
\begin{equation}\label{eq:PAULImatrices}
\B{I} = \begin{bmatrix} 1 & 0 \\ 0 & 1 \end{bmatrix}; \hspace{0.5em}
\B{X} = \begin{bmatrix} 0 & 1 \\ 1 & 0 \end{bmatrix}; \hspace{0.5em}
\B{Z} = \begin{bmatrix} 1 & 0 \\ 0 & -1 \end{bmatrix}.
\end{equation}
To represent arbitrary quantum operations also requires two-qubit gates, such as the well-known CNOT gate, which is the quantum version of the classical XOR gate:
\begin{equation} \B{CX}_{a,b} = \ket{0}\bra{0}\otimes\B{I} + \ket{1}\bra{1}\otimes\B{X}\label{eq:CX}\end{equation}
A CNOT gate flips the state of the target qubit $\ket{\B{b}}$ if and only if the control qubit $\ket{\B{a}}=\ket{\B{1}}$.
In fact, it can be shown that any multi-qubit logic gate (i.e., unitary matrix of arbitrary size) can be composed of CNOT and single-qubit gates, which is a concept called \textit{universality} of gate sets \citep{Barenco.1995}.
One universal set of gates, close to current quantum hardware, consists of CNOT \eqref{eq:CX}, X-Rotation \eqref{eq:RX}, and PhaseShift \eqref{eq:P} operations \citep{kitaev1997quantum}:
\begin{equation}\label{eq:RX}
\B{RX}(\lambda) = \exp\left(-i \frac{\lambda}{2} \B{X}\right)
\end{equation}
\begin{equation}\label{eq:P}
\B{P}(\phi) =  \exp\left(i\frac{\phi}{2}\right) \cdot \exp\left(-i\frac{\phi}{2} \B{Z}\right)
\end{equation}
In addition, we use the \textit{Controlled-PhaseShift} defined as:
\begin{equation}\label{eq:CP}
    \B{CP}(\phi) = \B{I} \otimes \ket{0} \bra{0} + \B{P}(\phi) \otimes \ket{1} \bra{1}
\end{equation}
Furthermore we refer to the gate set $\{\B{CX}, \B{H}, \B{S}\}$ as the \textit{Clifford group}, representing an alternative universal gate set with $\B{S} = \B{P}(\pi /2)$, and the Hadamard operator $\B{H}$: 
\begin{equation} \label{eq:hadamard}
 \B{H} = \frac{1}{\sqrt{2}}\begin{bmatrix} 1 & 1 \\ 1 & -1 \end{bmatrix} 
\end{equation}

\subsection{Reinforcement Learning}
To reflect the temporal component of executing operations in a quantum circuit, we formulate the problem of quantum circuit design as a \emph{Markov decision process (MDP)} $M = \langle \mathcal{S}, \mathcal{A}, \mathcal{P}, \mathcal{R}, \gamma \rangle$ with discrete time steps $t$, a set  $\mathcal{S}$ of states $s_t$, a set $\mathcal{A}$ of actions $a_t$, the transition probability $\mathcal{P}(s_{t+1} \mid s_{t},a_{t})$ from $s_t$ to $s_{t+1}$ when executing $a_t$, a scalar reward $r_t = \mathcal{R}(s_t,a_t)$, and the discount factor $\gamma \in [0,1)$ for calculating the \textit{discounted return}  \cite{puterman2014markov}: 
\begin{equation}
G_t = \sum^{\infty}_{k=0} \gamma^{t+k}r_{t+k}
\end{equation}
The goal is to find an optimal \emph{policy} $\pi^*$ with the action-selection probability $\pi(a_t \mid s_t)$ that maximizes the expected return.
Policy $\pi$ can be evaluated with the \textit{value function} $Q^{\pi}(s_t,a_t) = \mathbb{E}_{\pi}[G_t \mid s_t,a_t]$ or the \textit{state-value function} $V^{\pi}(s_t) = \mathbb{E}_{\pi,\mathcal{P}}[G_t \mid s_t]$. 
The \emph{optimal value function} $Q^{*} = Q^{\pi^{*}}$ of any \emph{optimal policy} $\pi^{*}$ satisfies $Q^{*} \geq Q^{\pi'}$ for all policies $\pi'$ and state-action pairs $\langle s_t, a_t \rangle \in \mathcal{S} \times \mathcal{A}$.

\emph{Value-based RL} approaches such as DQN approximate the optimal value function $Q^{*}$ using parameterized function approximators $Q_{\theta}$ to derive a behavior policy $\pi$ based on multiarmed bandit selection \cite{watkins1992q,mnih2015human}.
Alternatively, \emph{Policy-based RL} approximates $\pi^{*}$ from trajectories $\tau$ of experience tuples $\langle s_t, a_t, r_t, s_{t+1} \rangle$ generated by $\pi_{\theta}$, directly learning the policy parameters $\theta$ via gradient ascent on $\nabla_\theta \mathbb{E}_{\pi_\theta}\left[ G(\tau)\right]$ \cite{watkins1992q,sutton2018reinforcement}.  
For simplicity, we omit the parameter indices $\theta$ of $\pi$, $V$, and $Q$ for the following.
\textit{Advantageous Actor-Critic} (A2C) incorporates the value network $V^\pi$ to build the advantage $A_t^\pi=Q^\pi(s_t,a_t) - V^\pi(s_t)$, enabling updates using policy rollouts of incomplete episodes, where $\pi$ represents the actor and $V^\pi$ represents the critic \cite{mnih2016asynchronous}.
\textit{Proximal Policy Optimization} (PPO) offers an alternative approach, maximizing the surrogate objective 
\begin{equation}
L = \min\left\{\omega_t A_t^\pi, \text{clip}(\omega_t, 1 - \epsilon, 1 + \epsilon)A_t^\pi\right\},
\end{equation}
with $\omega_t = \frac{\pi_(a_{t} \mid s_{t})}{\pi_{\textit{old}}(a_{t} \mid s_{t})}$, where $\pi_{old}(a_{t} \mid s_{t})$ is the previous policy, and the clipping parameter $\epsilon$ restricts the magnitude of policy updates \cite{schulman2017proximal}.
PPO has been recently proven to ensure stable policy improvement, thus being a theoretically sound alternative to vanilla actor-critic approaches \cite{grudzien2022mirror}.
Similarly, applying clipping to twin Q functions in combination with delayed policy updates with \textit{Twin Delayed DDPG} (TD3) has been shown to improve \textit{Deep Deterministic Policy Gradient} (DDPG), approximating both $Q$ and $\pi$ \cite{dankwa2019twin}. 
Building upon those improvements, \textit{Soft Actor Critic} (SAC) bridges the gap between value-based Q-learning and policy gradient approaches and has shown state-of-the-art performance in various continuous control robotic benchmark tasks \cite{haarnoja2018soft}.

Yet, most RL algorithms face central challenges, such as exploring sparse reward landscapes and high-dimensional (continuous) action spaces. 
Nevertheless, besides robotic control or image-based games, RL has already been successfully applied to a wide range of current challenges in QC \cite{Bukov.2018,Colomer.2020,van2023qgym}.
By considering quantum circuit design as a sequential decision-making problem solved via RL, we hope to explore scalable yet adaptable approaches for their automated construction and optimization.
Note that alternative sequential decision-making approaches, like planning, might also be utilized.

\section{Quantum Circuit Design Objectives for Reinforcement Learning}\label{sec:objectives}
For a unified collection of RL challenges that arise from quantum architecture search and circuit optimization, we formalize generic \textit{state preparation} and \textit{unitary composition} objectives along with reward measures and specific target states in the following.

\subsection{State Preparation (SP)}
SP denotes the objective of finding a sequence of operations that will generate a desired quantum state from a given initial state $\ket{\B{0}}^{\otimes \eta}$, with the number of available qubits $\eta$. 
Denoting the final output state as $s_t = \ket{\B{\Phi}}\in\mathbb{C}^{2^\eta}$, the generated gate sequence can be seen as the unitary mapping
\begin{equation}\label{eq:unitarymapping}
\mathcal{U}: \ket{\B{0}}^{\otimes n}  \mapsto  \ket{\Phi}
\end{equation}
We measure success in this objective based on the distance between $\ket{\Phi}$ and the target state $\ket{\Psi}$ using the \textit{fidelity} $F$, measuring the overlap between the two states: 
\begin{equation}\label{eq:fidelity}
F(\Phi, \Psi) = \mid \braket{\Phi  \mid  \Psi} \mid^2 \quad \in\ [0,1]
\end{equation}
This definition intrinsically suggests a setting of sparse rewards for the SP objective, as we are not interested in the quantum state during an episode but rather want to maximize the final fidelity. 
% \todo{könnte man das nicht positiv verknüpfen? also wenn die distanz während der episode schon klein ist, dann reward geben, aber stets auch am ende der episode?}
Therefore, we define the accompanying reward as follows: 
\begin{equation}\label{eq:reward:statepreparation}
\mathcal{R}^{SP}(s_t,a_t) = F(s_t, \Psi)
\end{equation}
As a proof of concept, we propose using the following states: 
\begin{itemize}
\item The \textit{Bell state}, to reflect basic entanglement: 
\begin{equation}\label{eq:bellstate}    \ket{\Phi^+}\equiv\frac{\ket{00}+\ket{11}}{\sqrt{2}}
\end{equation}
\item The \textit{Greenberger–Horne–Zeilinger state} (GHZ), to reflect entanglement of a variable number of $n$ qubits:
\begin{equation}\label{eq:ghzstate}
\ket{GHZ} \equiv \frac{\ket{0}^{\otimes n}+\ket{1}^{\otimes n}}{\sqrt{2}}
\end{equation}
\item The \textit{Haar random state}, as a benchmark for preparing an arbitrary state:
\begin{equation}\label{eq:haarstate}
\ket{\Tilde{U}} =  \Tilde{U}\ket{\B{0}}^{\otimes \eta}
\end{equation}
\end{itemize}
However, note that further, more intricate states are easily extendable without any adaptations to the proposed framework.
Also, ignoring the global phase, this objective requires less accurate reproduction of the target, which we rectify by an alternative distance measure used for the following objective.

\subsection{Unitary Composition (UC)}
UC denotes the objective of composing an arbitrary unitary transformation with only a finite choice of operations.
UC can also be seen as a more general case of SP, in which the mapping of Eq.~\eqref{eq:unitarymapping} must not only apply to a specific input state but to all possible inputs. 
Note that while the choice of operations is finite, the gate set itself is arguably infinite due to including the parameter $\Phi$ into the action space. 
A necessary condition for decomposing a theoretical unitary circuit into a primitive set of operations is that this set is universal by means of \citet{Dawson2006}. 
The \textit{Solovay-Kitaev theorem}, however, states that there exists a finite sequence of gates that approximate any target unitary transformation $\B{U}$ to arbitrary accuracy without providing such a sequence.
Assuming that any circuit or sequence of gates $\Sigma_t = \langle a_0, \dots, a_t \rangle$, generated by policy $\pi$, can be represented as a unitary matrix $\B{V}(\Sigma_t)\in\mathbb{C}^{2^\eta\times 2^\eta}$, we define the distance $D$ to the target $V$ as: 
\begin{equation}
D(\B{U}, \B{V}(\Sigma_t)) = \norm{\B{U} - \B{V}(\Sigma_t)}^2,
\label{eq:distancebetweenmatrices}
\end{equation}
with $\norm{\cdot}$ denoting the Frobenius norm. 
As this can take on any positive real number and still reflects a loss to be minimized, we propose the following bounded \textit{similarity} score as a reward to be maximized for unitary composition:
\begin{equation}\label{eq:reward:unitarycomposition}
R^{UC}(a_t,s_t) = 1 - \arctan (D(\B{U}, \B{V}(\Sigma_t)))
\end{equation}
As a proof of concept, we use the following unitaries:
\begin{itemize}
\item The \textit{Hadamard} operator $\B{H}$ \eqref{eq:hadamard}, creating superposition
\item A random operator $\B{\Tilde{U}}$ according to \citet{ozols2009generate}
\item The \textit{Toffoli} operator defined as: 
\begin{equation}\label{eq:CCX}
CCX_{a,b,c} = \ket{0}\bra{0}\otimes\B{I}\otimes\B{I} + \ket{1}\bra{1}\otimes\B{CX},
\end{equation}
\end{itemize}

Lastly, we consider \textit{Hamiltonian simulation} (HS), which can be formalized as follows.
Given the Hamiltonian $\mathcal{H}$ of a quantum system, the solution to the Schrödinger equation yields a time evolution of the system according to $\ket{\Psi(\tau)} = \exp(-i\mathcal{H}\tau) \ket{\Psi_0}$. 
According to the Suzuki-Trotter formula, these dynamics can be approximately factorized with a rigorously upper-bounded error \citep{Lloyd.1996}. 
Thus, a sequence $\Sigma_t$ can be optimally chosen to produce
\begin{equation}\label{eq:trotterevolution}
\ket{\Phi} = \left ( \prod_t \Sigma_t\right ) \ket{\Psi},
\end{equation}
such that $\ket{\Phi}$ is as close as possible to $\ket{\Psi}$
 \cite{Bolens.2021}.
However, it can be observed that the simulation of the time evolution of a Hamiltonian $\mathcal{H}$ is mathematically equivalent to the UC of the unitary operator $\exp(-i\mathcal{H}\tau)$. 
While this fact can typically not be exploited in practice due to the required exponential computational resources in classical computing, we argue that RL should perform approximately equally in the UC and HS objectives due to their mathematical equivalence.

\section{Quantum Circuit Designer}\label{sec:designer}

To use RL for learning to solve the previously proposed objectives, we define \texttt{qcd-gym}, a generic \textit{quantum circuit design} (QCD) environment formalized as an MDP. % in the following
Fig.~\ref{fig:circuit-designer} shows a schematic rendering of the environment. 
To ensure maximal compatibility with various use cases and scenarios down the line, our framework only corresponds to a quantum device (simulator or real hardware), parameterized by its number of available qubits $\eta$ and its maximally feasible circuit depth $\delta$. 
Besides these parameters, used to constrain generated circuits to a practicable extent, we omit additional tunable parameters like solution thresholds.
To constantly monitor the current circuit, $\mathcal{P}(s_{t+1} \mid s_{t}, a_{t})$ is implemented by a quantum simulator, which allows for efficient state vector readout. 

\subsection{State}
At each time step $t$, the agent observes the state $s_t$ reflecting the current quantum circuit, represented by the full complex state vector $\ket{\Psi}$ or the unitary operator $\B{V}(\Sigma_t)$ resulting from the current sequence of operations $\Sigma_t$, and the intended target:
\begin{equation}\label{eq:state}
s_t = \begin{cases}
\langle\ket{\Psi},\ket{\Phi}\rangle\in\mathbb{C}^{2^{\eta+1}}\quad\quad \textit{target state }\ket{\Phi}\\
\langle\boldsymbol{V}(\Sigma_t), \boldsymbol{U} \rangle\in\mathbb{C}^{2^{2\eta+1}} \quad \textit{target unitary } \boldsymbol{U} 
\end{cases}
\end{equation}
This state representation is chosen as it contains the maximum amount of information about the output of the circuit.
Although this information is only available in quantum simulators (on real hardware, $\mathcal{O}(2^\eta)$ many measurements would be needed), it represents a starting point for RL from which future work should extract a sufficient and efficiently obtainable subset of information.
Furthermore, this state representation is sufficient for defining an MDP-compliant environment, as operations in this state must be reversible. 
Thus, no history of states is needed, and decisions can be made based solely on the current state.
Similar to continuous robotic control tasks \citep{gymnasium_robotics2023github}, we additionally provide the intended target to be observed.
Note that this addition does not require additional information, as the target representation is also necessary to calculate the reward. 
Furthermore, this representation can be extended to allow for mid-circuit measurements when switching to a density matrix-based representation of the state, however, at the cost of a quadratically larger state space.
All available qubits are initially in state $\ket{0}$.

\subsection{Action}
Given this observation, the agent needs to choose a suitable action consisting of three fragments: The operation $\Gamma$, the target $\Omega$, and the parameter $\Phi$.
To provide a balanced action space, we propose using an extension of the previously defined universal gate set, giving the agent a choice between controlled or uncontrolled operations along the $\mathbb{X}$ or $\mathbb{Z}$ axis. 
Additionally, the agent can choose to terminate the current episode, resulting in the following operation choice $\Gamma \in \{\mathbb{Z}, \mathbb{X}, \mathbb{T}\}$.
Upon episode termination, all unused qubits are disregarded.
Thus, the additional \textit{terminate} action $ \mathbb{T}$ enables the optimization of compact yet effective architectures within the defined limits for available qubits $\eta$ and feasible depth $\delta$.
If not terminated, an episode is truncated once the maximum circuit depth $\delta$ is reached. 
Secondly, the agent needs to choose the target of the operation, one qubit $q$ for applying the operation, and one control qubit $c$, $\Omega = \{q,c\} \in [0,\eta-1]^2$. 
For the \textit{terminate} operation, the target fragment is discarded. 
If the operation qubit equals the control qubit, an uncontrolled operation is performed, resulting in the following set of gates $\mathbb{G}=\{\B{RX}, \B{CX}, \B{P}, \B{CP} \}$ (cf. Eqs.~\eqref{eq:CX}-\eqref{eq:CP}) to be applied: 
\begin{equation} \label{eq:gate}
g(o,q,c): \Gamma \times \Omega \mapsto \mathbb{G} = 
\begin{dcases} \begin{dcases} 
\B{RX} & q = c \\ \B{CX} & q \neq c 
\end{dcases} & o = \mathbb{X} \\  \begin{dcases} 
\B{P} & q = c \\ \B{CP} & q \neq c 
\end{dcases} & o = \mathbb{Z} \\  \end{dcases}
\end{equation}
Selecting actions from a universal set of gates, based on the current state, ensures the production of only valid circuits. 
In contrast to most related work, we refrain from using a discrete action space with a gate set like \textit{Clifford + T}.
Instead, we aim to exploit the capabilities of RL to perform continuous control over the gate parameterization.
Thereby, we enable granular state-control via the last action fragment $\Phi \in [-\pi,\pi]$, facilitating the generation of optimized circuits in a unified circuit design approach.
Otherwise, a second optimization pass would be needed to improve the placed continuous gates, increasing the application's overall complexity. 
Also, using an external optimizer for parameterization does not suit our approach, as we aim to benchmark the suitability of RL for the design of optimized quantum circuits.
Note that the parameter is omitted for the CNOT gate $\B{CX}$.
This helps reduce the action space and is motivated by the empirical lack of CRX gates in most circuit architectures.
Overall, this results in the $4$-dimensional action space $\langle o_t, q_t, c_t, \Phi_t \rangle = a_t \in  \mathcal{A} = \{\Gamma \times \Omega \times \Theta\}$.
Thus, following Eq.~\eqref{eq:gate}, the operation $U = g(o_t,q_t,c_t)(\Phi_t)$ is applied at time step $t$.

\subsection{Reward}
The reward is kept at $0$ until the end of an episode is reached (i.e., $t$ is terminal, either by truncation or termination). 
To incentivize the use of few operations, a step-cost $\mathcal{C}_t$ is applied when exceeding two-thirds of the available operations $\sigma$:
\begin{equation}\label{eq:stepcost}
\mathcal{C}_t=\max\left(0,\frac{3}{2\sigma}\left(t-\frac{\sigma}{3}\right)\right), 
\end{equation}
with $\sigma = \eta \cdot \delta \cdot 2$.
We choose this bound to prevent prevalent solutions solely optimizing for minimal qubit and operation usage, which would cause immediate termination of the episode. 
To further incentivize the construction of useful circuits, we use the previously defined task reward functions $\mathcal{R}^{*}$ according to Eq.~\eqref{eq:reward:statepreparation} and Eq.~\eqref{eq:reward:unitarycomposition} for the SP and UC objectives respectively, s.t.:
\begin{equation}\label{eq:reward}
\mathcal{R} = \begin{dcases} \mathcal{R}^{*}(s_t,a_t)-C_t, &  \text{if $t$ is terminal.}  \\0, & \text{otherwise.} \end{dcases}
\end{equation}
However, in contrast to the overall domain, those are specific to the given problem instance or objective, defined previously, and might be extended to address further objectives.

\section{Related Work}

\subsection{Quantum Architecture Search}
Currently, prevailing quantum circuit architectures are either manually designed -- encoding problem-specific domain-knowledge in, e.g., \textit{variational quantum eigensolvers} (VQEs) \citep{Peruzzo.2014} and approximate methods like the \textit{quantum approximate optimization algorithm} model (QAOA) \citep{farhi2014quantum}  --  or heuristically constructed via iterative search. In this work, we specifically consider quantum architecture search for the composition of unitary gate circuit (UC), the preparation of quantum states (SP), and the simulation of a given Hamiltonian (HS) \citep{Bharti.2022}. \textit{Quantum architecture search} (QAS) denotes the process of automatically engineering quantum circuits to find a sequence of quantum gates achieving specific tasks. Due to the likeness to the similarly connected and layered classical (neural-)network structures, QAS, in some cases, also takes inspiration from neural architecture search (NAS) \cite{zoph2016neural}. Hence, deep-learning NAS concepts like augmented~\cite{Selig2023DeepQPrep} or differentiable architecture search \citep{Zhang.2022} and topological (graph-)augmentation \cite{stanley2002evolving} have found application in QAS in similar form \cite{giovagnoli2023qneat}.

Recently, evolutionary-inspired algorithms are also commonly used for their effectiveness in searching large problem spaces \cite{rattew2019domain, chivilikhin2020mog, sunkel2023ga4qco, ding2008evolving}. However, their convergence to a few best solutions out of a preset population of candidates does imply the generation of finely specialized, problem-specific circuits.

In contrast, motivated by the sequential nature of quantum circuits, we approach QAS as a dynamic decision-making problem. Therefore, we optimize a policy to select actions consisting of a quantum gate placement such that the final circuit maximizes the cumulative reward for a given objective. Utilizing RL/ML for this purpose has been shown to work with both the optimization and the generation of quantum circuits. \citet{ostaszewski2021reinforcement}, for instance, propose using RL to iteratively generate VQE Hamiltonians, or more generally, \citet{mortazavi2022theta} propose a single-step MDP for the exploration of optimization sample-choices. However, their work denotes a problem-specific top-down approach, whereas we strive for a bottom-up approach, formulating generic RL challenges from the field of QAS.
% UC
Likewise, RL approaches have been proposed to prepare certain quantum states of varying difficulty \cite{kuo2021quantum, Sogabe.2022,gabor2022applicability}, producing a more generic challenge.
Similar to \citet{Sogabe.2022}, we propose a reward based on the distance between the final and target states, where \citet{kuo2021quantum} use a distance threshold, introducing an additional hyperparameter. 

To the best of our knowledge, existing approaches essentially utilize discrete gate sets, such as the Clifford group defined above, e.g., \cite{koelle2024reinforcement, ostaszewski2021reinforcement}.
Using operations with predefined parameterizations might require a two-step procedure: generating a circuit architecture and then optimizing the circuit parameterization. Classical optimization routines are traditionally employed to perform this task, including gradient-free approaches, such as evolutionary algorithms, and gradient-based approaches using the parameter-shift rule \cite{crooks2019gradients}. Note that both methods require numerous circuit executions to find optimal parameters. In contrast, our approach seeks to integrate this parameter optimization task directly into the RL environment by enabling continuous actions. Thus, agents inherently seek optimal parameters in the generated circuits.

\subsection{Quantum Control}
In this paper, we propose the challenge of composing a target unitary rather than performing a backward search starting from the target state since the decomposition of prepared states or already given unitary operations constitutes a well-known challenge \cite{Zhang2020, Pirhooshyaran.2021, mansky2023near}. Composing such a valid unitary circuit for specific input states and all possible inputs requires high-precision operations. Since the tasks considered in this work (state preparation and unitary approximation) are commonly considered in quantum control, we also briefly highlight two related approaches that might constitute suitable baseline candidates but fall outside the scope of our proposed RL structure.

% GRAPE 
The GRAPE (Gradient Ascent Pulse Engineering) algorithm is used in quantum control to find optimal control pulses. The essence of GRAPE is to maximize a target function that represents the fidelity of the quantum operation being controlled relative to a desired quantum operation. It does this by iteratively adjusting the control pulses based on the gradient of the fidelity concerning the controls. This approach helps fine-tune quantum operations to achieve higher precision and efficiency in quantum computing systems.\cite{de2011second}

%  CRAB
The CRAB (Chopped Random Basis Quantum Optimization) algorithm is designed for quantum optimal control. It works by expressing the control fields as expansions on a randomly chosen basis, which simplifies the problem from optimizing a functional form to finding optimal parameters on this basis. The algorithm iteratively adjusts these parameters to maximize the fidelity of the quantum operation. This method is particularly efficient because it reduces the complexity of the control problem, allowing for faster and more efficient optimization in quantum systems. You can read more about it in the full article here. \cite{caneva2011chopped}

\subsection{Quantum Circuit Optimization}
In addition to searching for feasible circuit architectures, RL has also been considered for \textit{quantum circuit optimization} (QCO). 
% Structure Optimization
The most apparent challenge in QCO is optimizing the circuit structure to reduce its depth and global gate count. 
The goal is to find a logically equivalent circuit using fewer gates from a given quantum circuit, thus reducing the runtime. 
In addition to approaches such as the use of ZX calculus \citep{Kissinger.2020}, employing RL has also been proposed, either, with the agent operating on the entire circuit \citep{Fosel.3132021}, or, incorporating optimization through a gate-by-gate procedure \citep{ostaszewski2021reinforcement}. 
\citet{ostaszewski2021reinforcement} even argue that RL will naturally find shorter and better-performing solutions, as it is designed to maximize the discounted sum of rewards. 
In practice, we propose applying a step cost for each executed operation.
Therefore, the structural optimization of the circuits is inherent in the design of our approach. 

% Accommodate Hardware limitations 
To achieve practical effectiveness, quantum circuits must also be designed to accommodate the constraints of existing NISQ hardware. Beyond the circuit depth, these constraints include the limited number and connectivity of qubits, the available gate set, and the high susceptibility to errors \citep{Bharti.2022}.
% Qubit Routing Problem / Connectivity
The qubit routing problem describes the challenge of mapping logical to physical qubits \citep{Childs.2019}, which might not be connected due to experimental limitations. 
SWAP operations are applied to connect pairs, allowing arbitrarily placing two-qubit gates, such as CNOT.
However, taking the resulting circuit depth into account, this constitutes an NP-hard combinatorial optimization problem \citep{Siraichi.2018}. 
Besides popular approaches such as sorting networks \citep{Steiger.2018}, or proprietary compiler-specific methods, the application of RL operating on the entire circuit has been shown to outperform state-of-the-art methods \cite{Herbert.30.12.2018,Pozzi.2020}.
% Error Mitigation 
Regarding the design of high-fidelity error-tolerant gate sets, compensating for the processes that induce hardware errors on NISQ devices, \citet{Baum.2021} proposed using RL to manipulate a small set of accessible experimental control signals.
Considering the qubit routing problem and error mitigation, our approach does not yet consider these hardware-specific properties, but it can be easily extended to do so. 
Concerning reduced hardware connectivity, the action space for multi-qubit gates could be reduced to qubits that are directly connected to each other in the hardware topology. 
Regarding Error Mitigation, the state of the RL agent could be changed from state vectors to density matrices to model errors occurring, e.g., due to decoherence or imperfect gates.

Overall, those applications further underscore the efficacy of RL in quantum computing. 
Yet, most approaches consider RL for specific applications, using specialized discrete gate sets, where the parameters need to be optimized post hoc.

\section{Implementation}
The QCD environment is implemented based on \textit{gymnasium} \citep{gymnasium} and \textit{qiskit} \citep{qiskit}.
All implementations are available here
\footnote{\url{https://github.com/philippaltmann/QCD}}.
% \footnote{\url{https://pypi.org/p/qcd-gym}}.

\subsection{Environments}
As previously elaborated, state preparation (SP) and unitary composition (UC) objectives are either closely related to the others mathematically, accounted for by the environment design, or can be easily incorporated. 
As defined in \autoref{sec:objectives}, we used states $\Psi \in \{ \ket{\Phi^+}, \ket{\Tilde{U}},  \ket{GHZ}\}$, with $\eta\in \{2,2,3\}$ and $\delta \in \{12,12,15\}$ respectively, for SP, and unitaries $U\in\{\B{H}, \B{\Tilde{U}}, \B{CCX}\}$, with $\eta\in \{1,2,3\}$ and $\delta \in \{9,12,63\}$ respectively, for UC.
As a proof-of-concept for the proposed environment architecture, we only conducted experiments for this initial set of targets. 
However, note that the QCD allows for easy extension of both objectives by defining arbitrary target states and unitaries. 

\subsection{Action Space}
Regarding the implementation of the action space, the main challenge we face is the mixed nature of quantum control: 
While selecting gates and qubits to be applied resembles a discrete decision problem, we argue that integrating continuous control over the parameterization of the placed gate is crucial to unveiling RL's full potential for QC. 
As current RL algorithms do not implement policies with mixed output spaces, we opt for a fully continuous action space and discretize the choice over the gate, qubit, and control contained.
Yet, our long-term vision is to enable mixed control to suit quantum control optimally.

\subsection{Baselines}
To provide a diverse set of state-of-the-art model-free baselines, we decided for
A2C \cite{mnih2016asynchronous}, PPO \cite{schulman2017proximal},  TD3 \cite{dankwa2019twin}, and SAC \cite{haarnoja2018soft}.
A discretization of the proposed action space (especially the parameterization) would require $3\cdot\eta^2\cdot\varrho$ dimensions, with a resolution $\varrho=4$, resulting in more than 100 dimensions of actions only for a 3-qubit circuit.
We, therefore, omitted DQN baselines, which only support discrete action spaces.
All baselines are implemented extending \citet{stable-baselines3} and use the default hyperparameters suggested in their corresponding publication. 
All policies are trained for 1M time steps.
Additionally, we provide a \textit{Random} baseline averaged over 100 episodes of randomly chosen actions.
To reflect alternative state-of-the-art approaches for automated quantum circuit design, we furthermore provide results from a genetic algorithm (\textit{GA}).
The approach is implemented according to \citet{sunkel2023ga4qco}, with 20 individuals optimized for 50 epochs, using the cost-adjusted objectives previously defined to evaluate their fitness. 
Furthermore, note that we chose the maximum depth $\delta$ for all non-random tasks such that an optimal handcrafted solution would require only a third of the available depth using all available qubits, which indirectly serves as an additional human baseline.
For significance, all results are averaged over eight random seeds, reporting both the mean and the 95\% confidence interval. 

\subsection{Metrics}
As a performance metric, we report the pure objective rewards for the final circuits (without applying the operation cost): 
The \textit{Mean Fidelity} according to Eq.~\eqref{eq:reward:statepreparation} for state preparation tasks and the \textit{Mean Similarity} according to Eq.~\eqref{eq:reward:unitarycomposition} for unitary composition tasks. 
To reflect capabilities regarding circuit optimization or the generation of optimized circuits, we furthermore provide the final number of utilized qubits (\textit{Mean Qubits}) and the final depth of the circuit (\textit{Mean Depth}). 
Note that the final depth of the circuit is affected by the scheduling of the sequence of operations generated by the policy, which is currently handled by the simulation. 
However, scheduling tasks have also been shown to be solvable RL objectives and could be integrated into the reward function to reduce the need for external compilation.
Note that, regardless of defining constraints for the number of qubits $\eta$ and the depth $\delta$ beforehand, the choice on the number of operations is made by the policy. 
For smoothing, all metrics are averaged over a sliding window of 100 episodes (or individuals for the GA).

\begin{figure*}[t]\centering
\includegraphics[width=.5\linewidth]{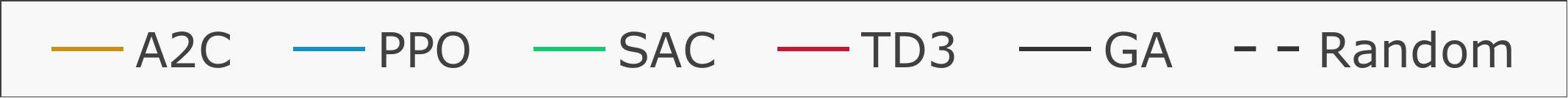}
  \subfloat[Hadamard Composition $\mid$ Mean Similarity]{\includegraphics[width=.32\linewidth]{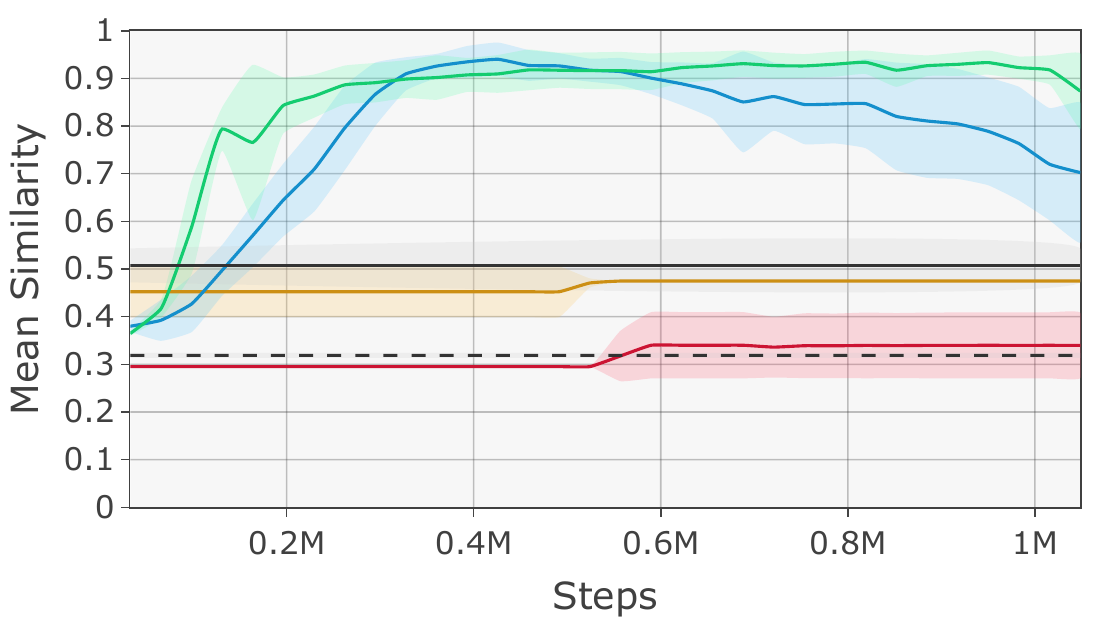} \label{fig:eval:hadamard-return}}
  \subfloat[Hadamard Composition $\mid$ Mean Qubits]{\includegraphics[width=.32\linewidth]{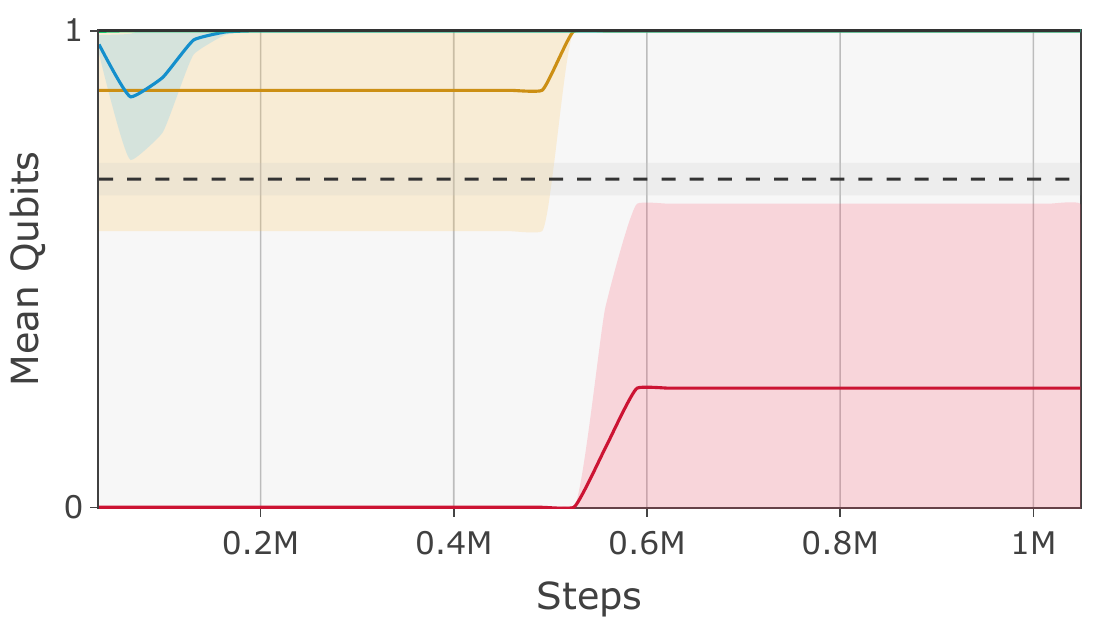} \label{fig:eval:hadamard-qubits}}
  \subfloat[Hadamard Composition $\mid$ Mean Depth]{\includegraphics[width=.32\linewidth]{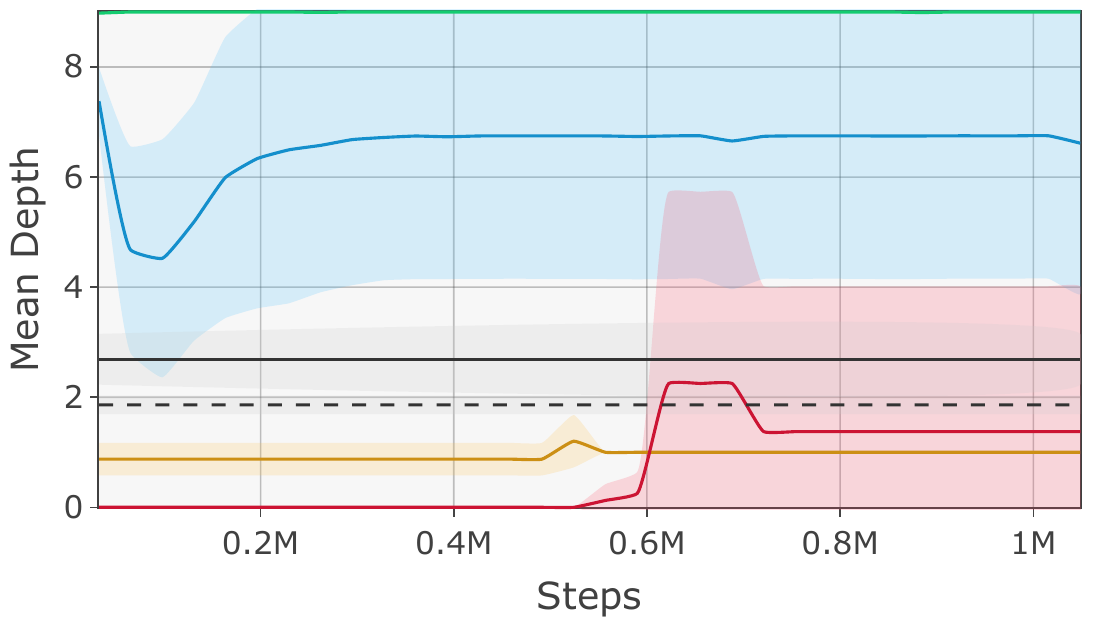} \label{fig:eval:hadamard-depth}}\\
/  \subfloat[GHZ State Preparation $\mid$ Mean Fidelity]{\includegraphics[width=.32\linewidth]{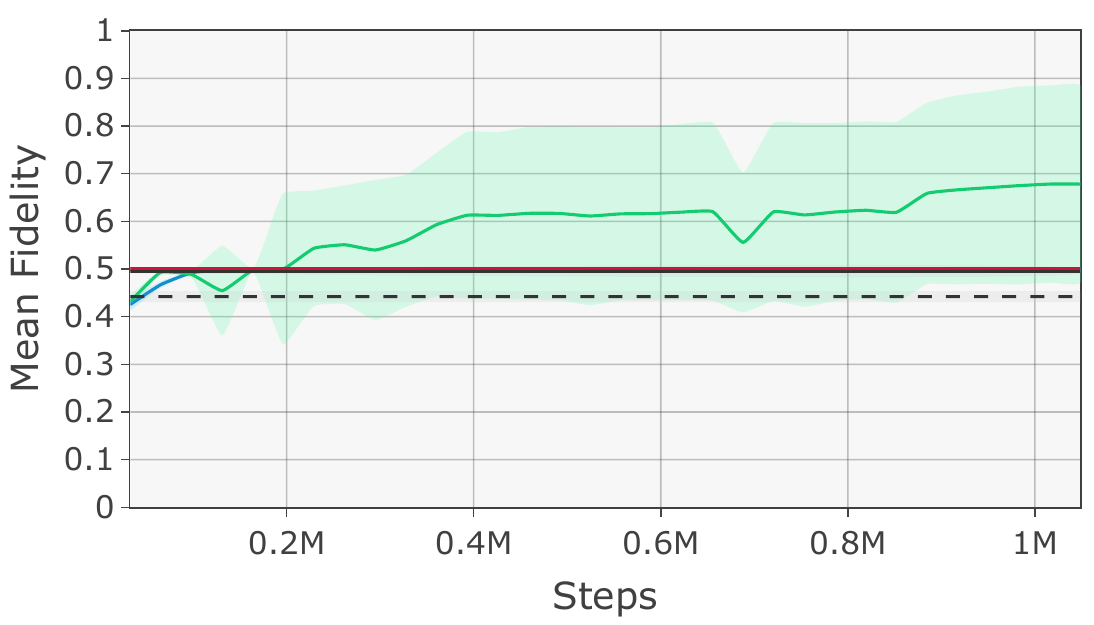} \label{fig:eval:ghz-return}}
  \subfloat[GHZ State Preparation $\mid$ Mean Qubits]{\includegraphics[width=.32\linewidth]{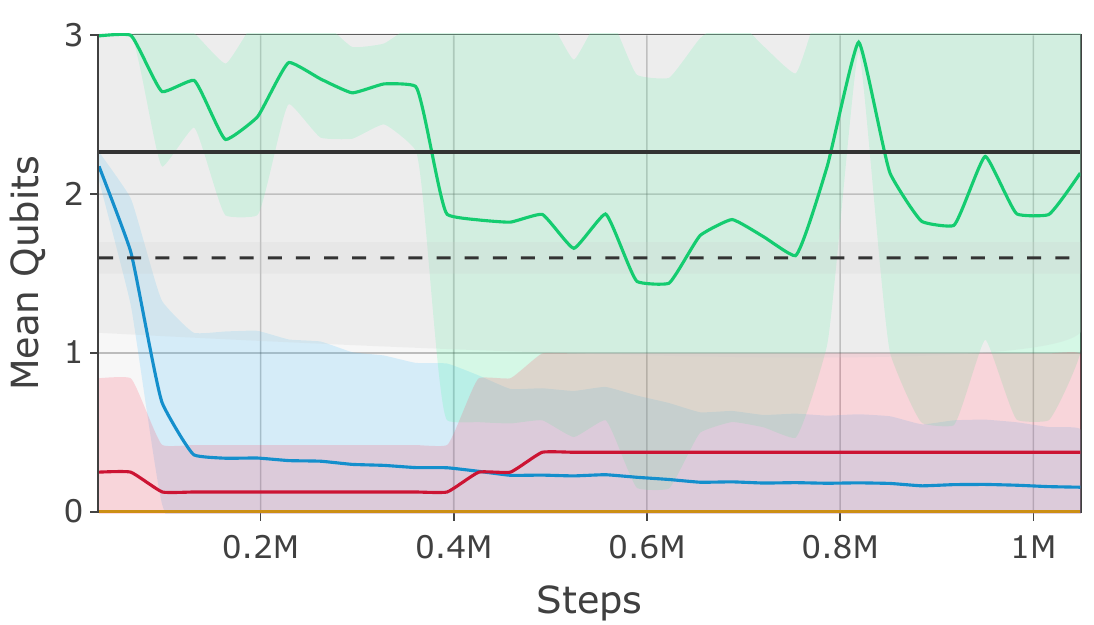} \label{fig:eval:ghz-qubits}}
  \subfloat[GHZ State Preparation $\mid$ Mean Depth]{\includegraphics[width=.32\linewidth]{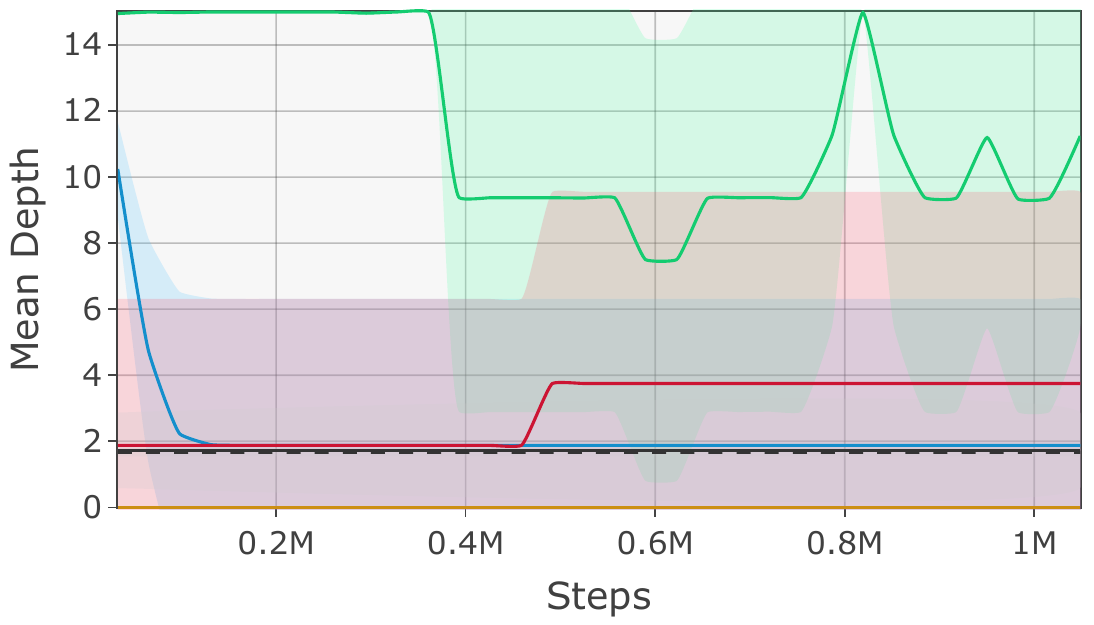} \label{fig:eval:ghz-depth}}\\
\caption{\textbf{Quantum Circuit Design Evaluation:} Benchmarking A2C (orange), PPO (blue), SAC (green), and TD3 (red) for Hadamard Composition (Fig.~\ref{fig:eval:hadamard-return}-\ref{fig:eval:hadamard-depth}) and GHZ State Preparation (Fig.~\ref{fig:eval:ghz-qubits}-\ref{fig:eval:ghz-depth}) with regards to the Mean Metric (Fidelity and Similarity, higher is better), Mean Qubits utilized, and Mean Depth of the resulting circuit, against a GA (gray) and a Random baseline (dashed line). Shaded areas mark the 95\% confidence intervals. Overall, SAC shows the highest objective performance with the highest qubit and depth utilization (which could be further improved towards the optimal 1/3 operation utilization).} \label{fig:eval:1}
\end{figure*}
    
\begin{figure*}[t]\centering
\includegraphics[width=.5\linewidth]{assets/Legend.pdf}
\subfloat[Random SP $\mid$ Mean Fidelity]{\includegraphics[width=.32\linewidth]{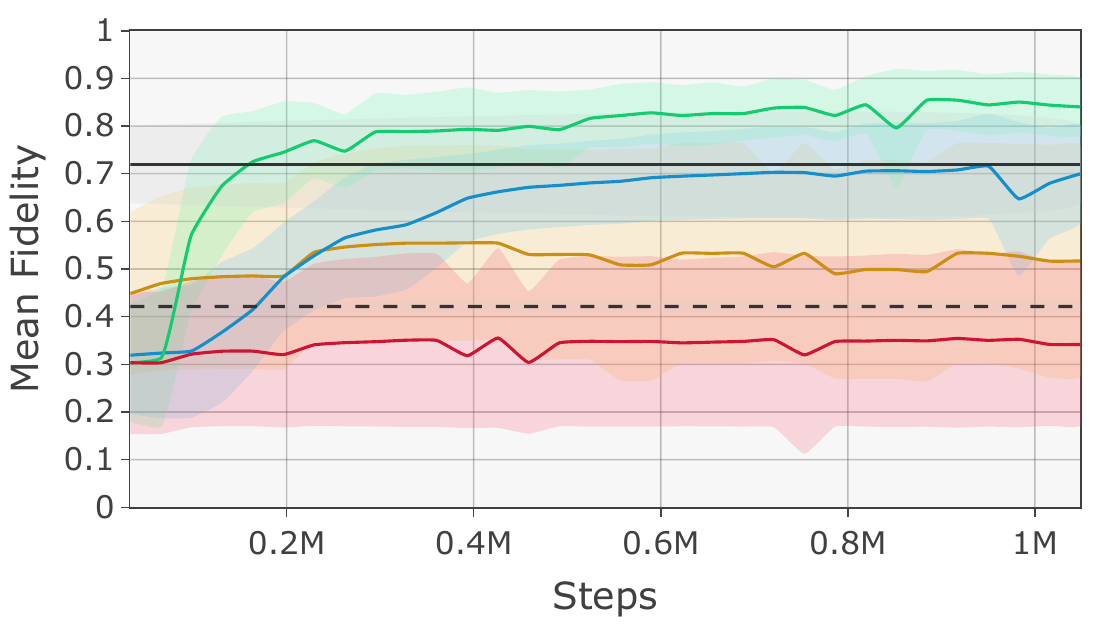} \label{fig:eval:random-state-return}}
\subfloat[Random SP $\mid$ Mean Qubits]{\includegraphics[width=.32\linewidth]{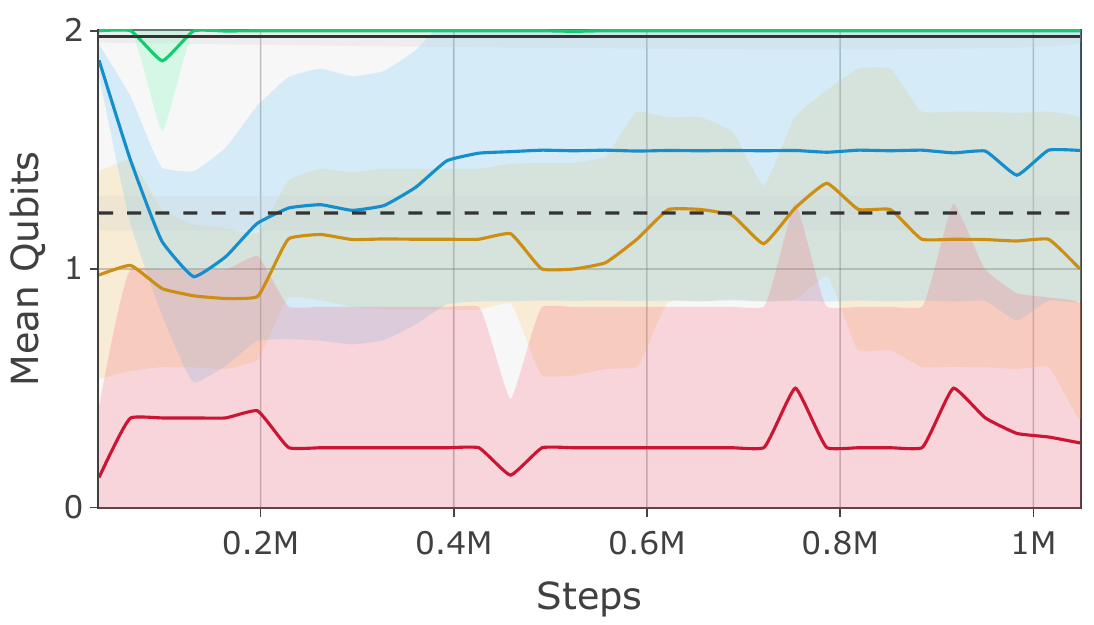} \label{fig:eval:random-state-qubits}}
\subfloat[Random SP $\mid$ Mean Depth]{\includegraphics[width=.32\linewidth]{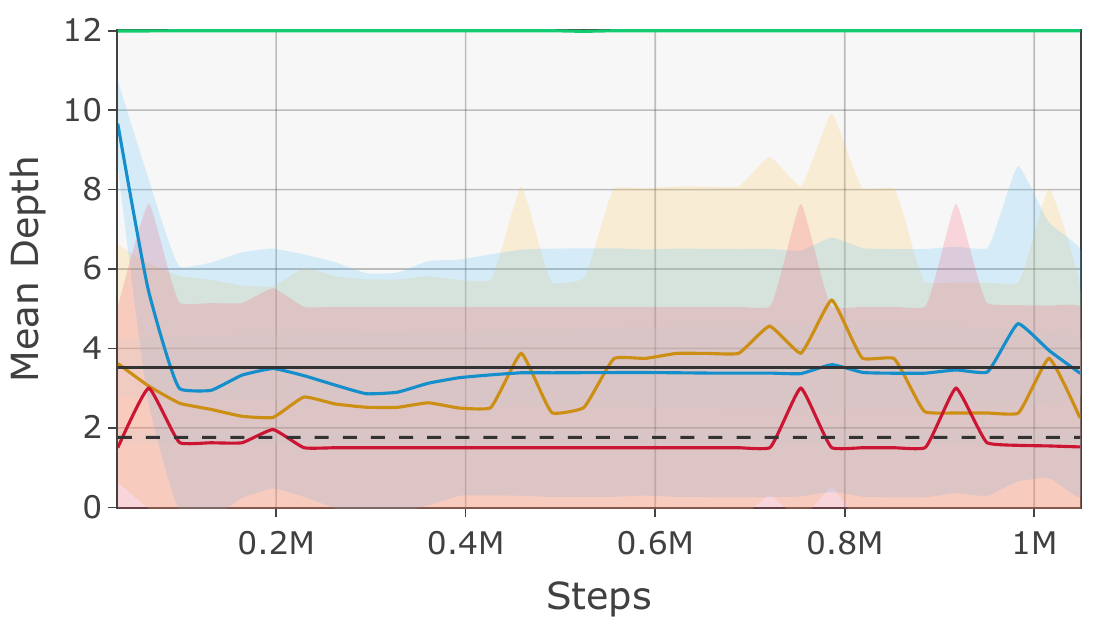} \label{fig:eval:random-state-depth}}\\  
\subfloat[Toffoli Composition $\mid$ Mean Similarity]{\includegraphics[width=.32\linewidth]{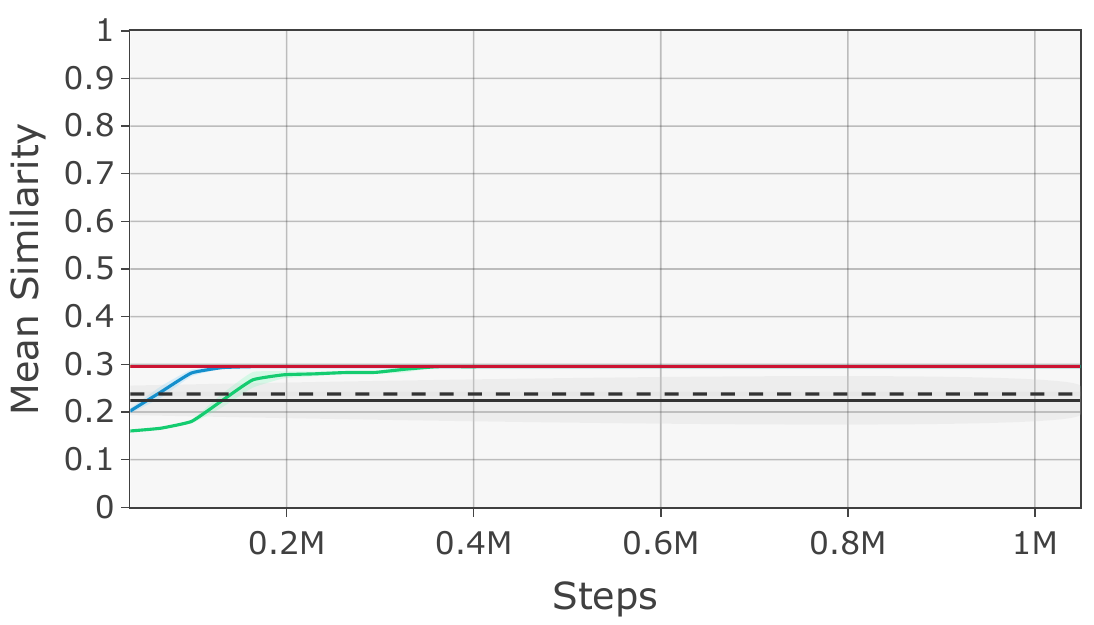} \label{fig:eval:toffoli-return}}
\subfloat[Toffoli Composition $\mid$ Mean Qubits]{\includegraphics[width=.32\linewidth]{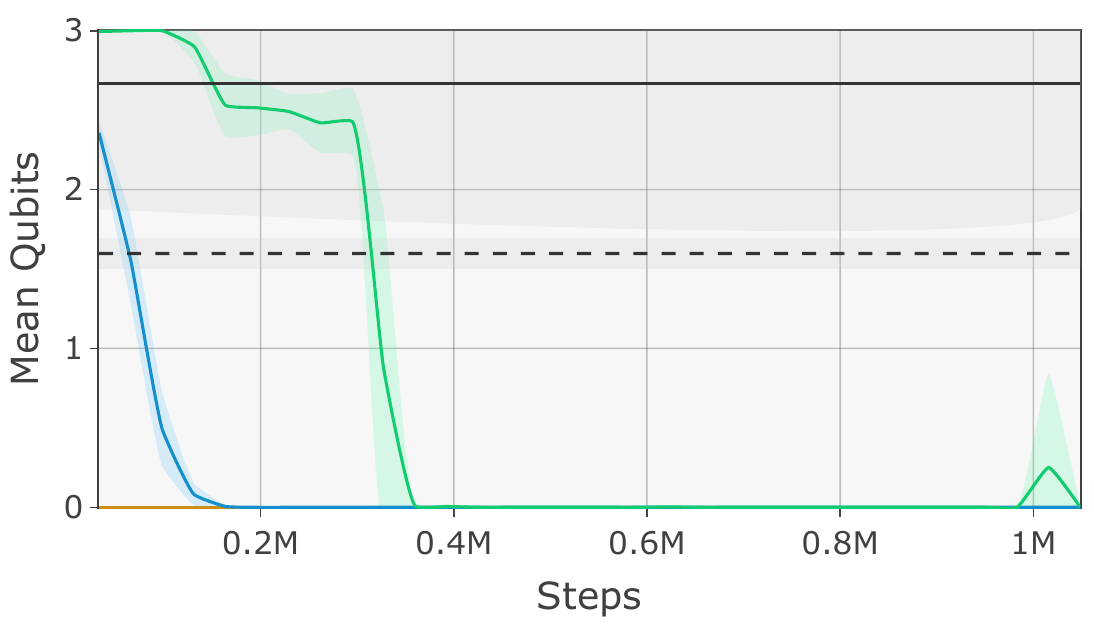} \label{fig:eval:toffoli-qubits}}
\subfloat[Toffoli Composition $\mid$ Mean Depth]{\includegraphics[width=.32\linewidth]{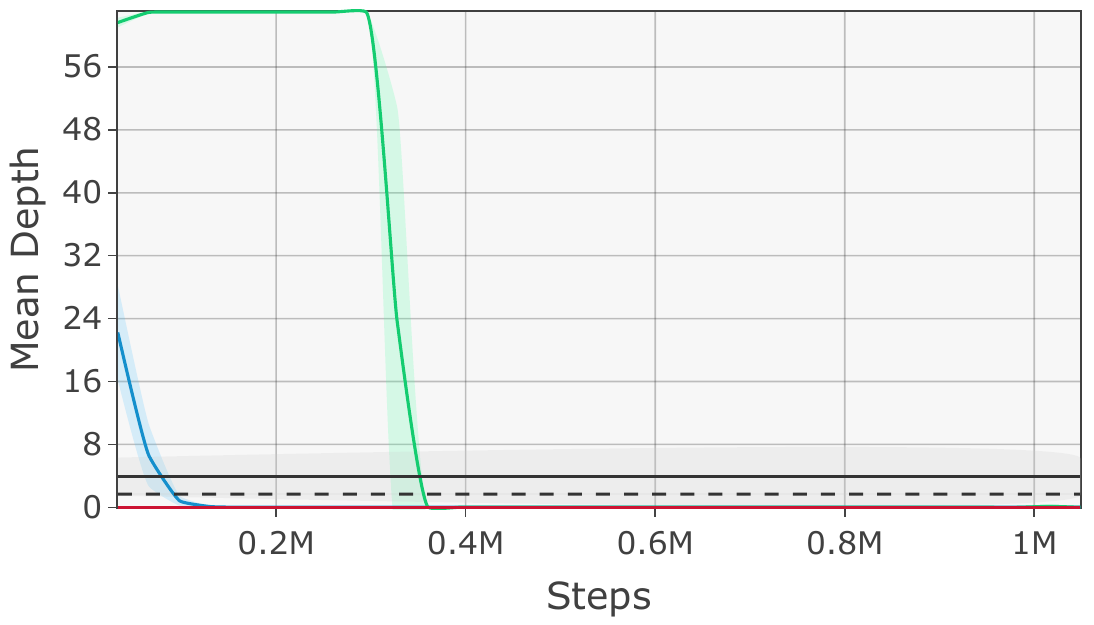} \label{fig:eval:toffoli-depth}}\\
\caption{\textbf{Quantum Circuit Design Evaluation:} Benchmarking A2C (orange), PPO (blue), SAC (green), and TD3 (red) for Random State Preparation (Fig.~\ref{fig:eval:random-state-return}-\ref{fig:eval:random-state-depth}) and Toffoli Composition (Fig.~\ref{fig:eval:toffoli-return}-\ref{fig:eval:toffoli-depth}) with regards to the Mean Metric (Fidelity and Similarity, higher is better), Mean Qubits utilized, and Mean Depth of the resulting circuit, against a GA (gray) and a Random baseline (dashed line). Shaded areas mark the 95\% confidence intervals. While random states are prepared predominantly well, utilizing all qubits and reasonable depths, the Toffoli Composition exhibits a local similarity optimum of 0.3 for empty circuits.}\label{fig:eval:2}
\end{figure*}

\section{Evaluation}
In the following, we aim to demonstrate the challenges of current RL approaches inherent in QCD. 
First, we look at two elementary objectives:
Constructing the Hadamard operation and preparing the 3-qubit GHZ state.
As introduced earlier, the Hadamard operator is an integral component for exploiting superposition in QC and is, therefore, included in most elementary gate sets. 
However, as we focus on the choice of parameterized gates, it is precluded from our action space. 
Playing a central role in most quantum algorithms nevertheless, it is crucial for RL to reconstruct it.
Reconstruction is possible via $H = P(\pi/2) RX(\pi/2) P(\pi/2)$. 
Connecting the Hadamard state to another or multiple qubits via one or multiple CNOTs yields the Bell and GHZ states, respectively. 
Both states likewise play a vital role in QC algorithms, expressing maximum entanglement of the involved qubits. 
Given their close resemblance, please refer to the Appendix for further results for preparing the Bell state.

Fig.~\ref{fig:eval:1} shows the evaluation results.
Regarding the Mean Similarity for the Hadamard composition, all compared algorithms exceed the random baseline. 
Yet, presumably due to the intricate action space, A2C and TD3 stagnate at a performance below the GA at $0.5$. 
PPO and SAC, on the other hand, exceed this local optimum for nearly empty circuits and reach an above-$90\%$ resemblance to the target Hadamard unitary within 400k time steps.
Looking at the qubit and depth utilization reveals that both approaches converge to steadily using the available qubit. 
While PPO converges to a mean depth slightly above six, SAC utilizes the full available depth, which indicates further potential for optimizing toward the secondary compactness objective. 
Overall, because we intend to build a single framework for designing optimized circuits, QCD is prone to getting stuck in local minima, which is introduced by its nature in multi-objective optimization. 

\begin{figure*}[h]\centering
\includegraphics[width=.5\linewidth]{assets/Legend.pdf}\\
\subfloat[Bell State Preparation $\mid$ Mean Return]{\includegraphics[width=.32\linewidth]{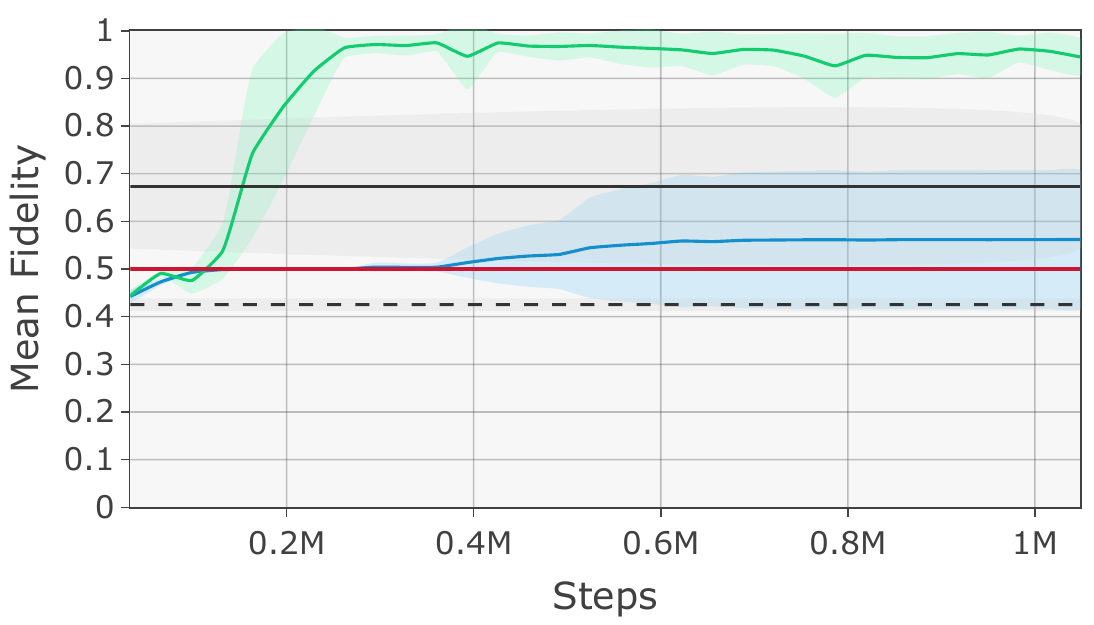} \label{fig:eval:bell-return}}
\subfloat[Bell State Preparation $\mid$ Mean Qubits]{\includegraphics[width=.32\linewidth]{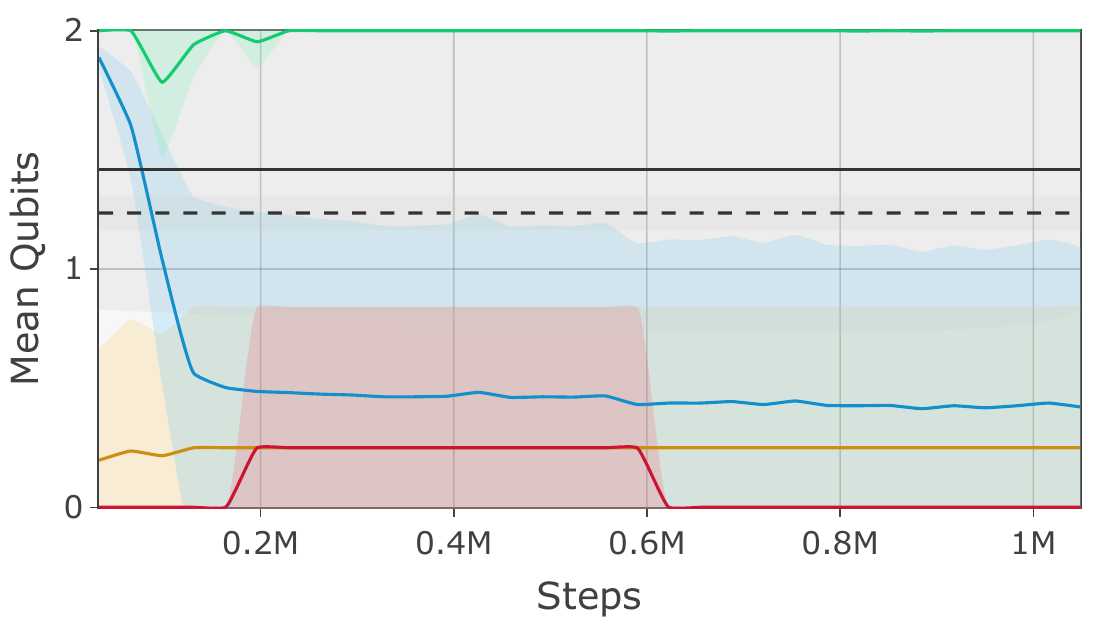} \label{fig:eval:bell-qubits}}
\subfloat[Bell State Preparation $\mid$ Mean Depth]{\includegraphics[width=.32\linewidth]{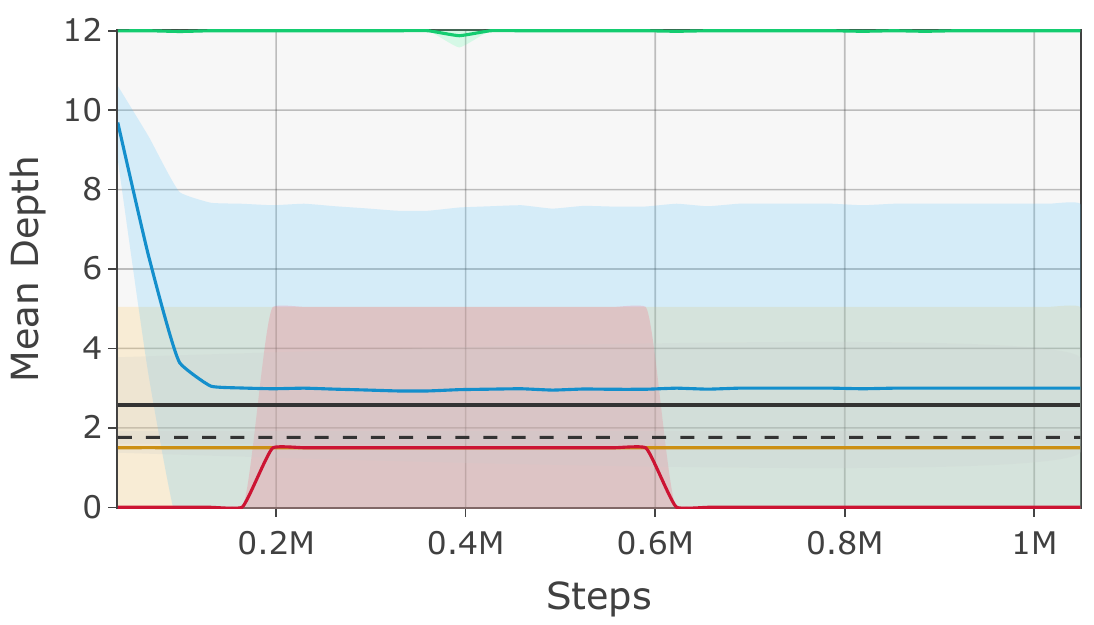} \label{fig:eval:bell-depth}}\\
\subfloat[Random Composition $\mid$ Mean Return]{\includegraphics[width=.32\linewidth]{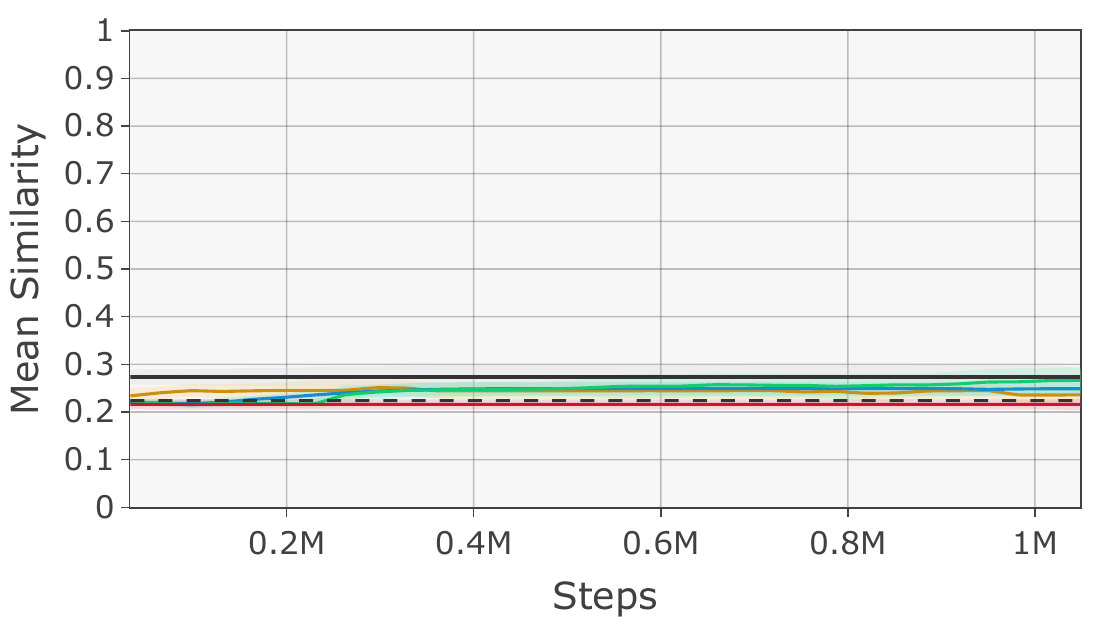} \label{fig:eval:random-composition-return}}
\subfloat[Random Composition $\mid$ Mean Qubits]{\includegraphics[width=.32\linewidth]{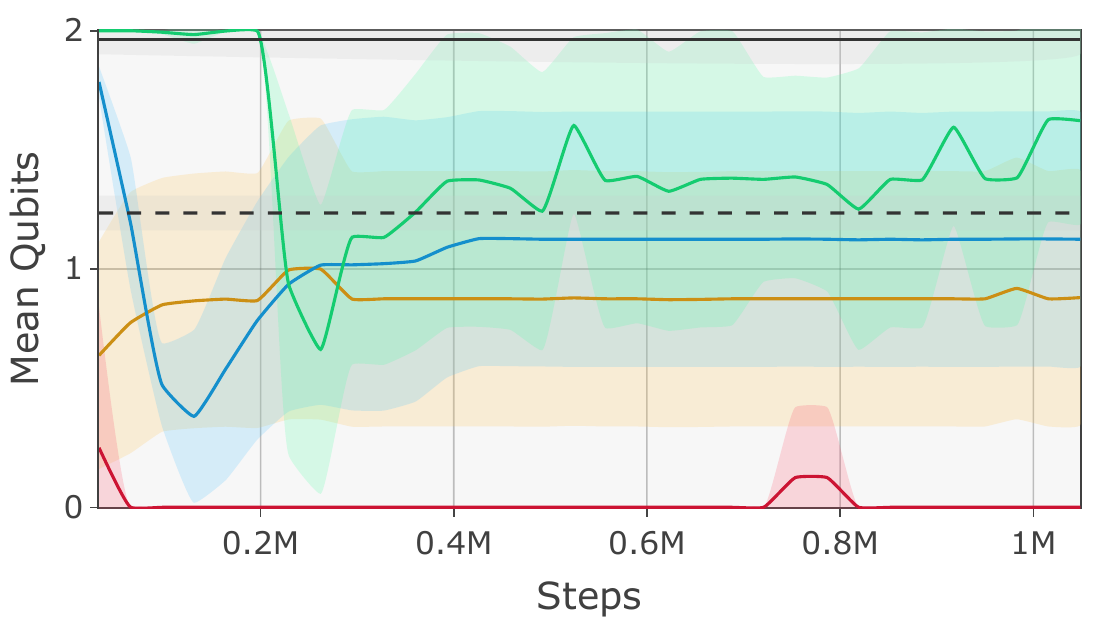} \label{fig:eval:random-composition-qubits}}
\subfloat[Random Composition $\mid$ Mean Depth]{\includegraphics[width=.32\linewidth]{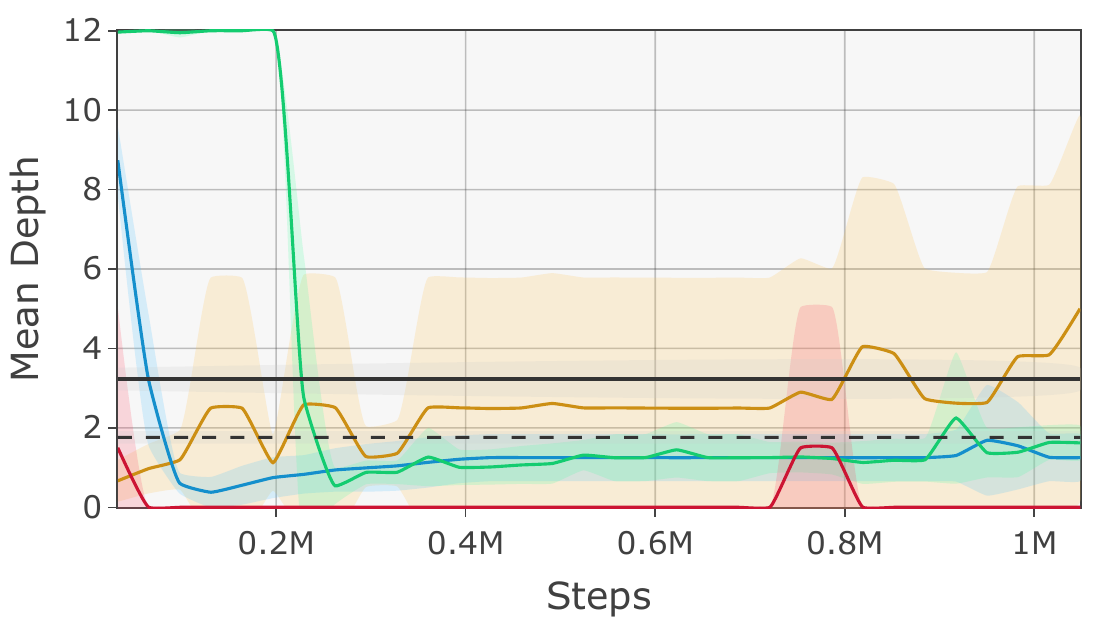} \label{fig:eval:random-composition-depth}}\\  
\caption{\textbf{Additional Benchmark Results:} Benchmarking A2C (orange), PPO (blue), SAC (green), and TD3 (red) for Bell State Preparation (Fig.~\ref{fig:eval:bell-return}-\ref{fig:eval:bell-depth}) and Random Composition (Fig.~\ref{fig:eval:random-composition-return}-\ref{fig:eval:random-composition-depth}) with regards to the Mean Metric (Fidelity and Similarity, higher is better), Mean Qubits utilized, and Mean Depth of the resulting circuit, against a GA (gray) and a Random baseline (dashed line). Shaded areas mark the 95\% confidence intervals. While the Bell state is almost optimally prepared using SAC, composing random unitary operations yields serious challenges that hinder optimal convergence.} \label{fig:eval:3}
\end{figure*}

Given that the subsequent task of creating entanglement builds upon the skill of putting qubits into superposition, only SAC and PPO can be expected to succeed in this task. 
This constitutes a further challenge that QC yields for RL: 
Complex circuit architectures are often built upon hierarchical building blocks.
This characteristic justifies multi-level optimization approaches resembling similar hierarchies. 
On the other hand, the proposed low level might yield the exploration of unconventional approaches providing out-of-the-box solutions to complex multi-level challenges. 
Indications of said prospects can be observed for the GHZ state preparation results, where SAC reaches a mean fidelity of around $0.7$, again significantly outperforming the compared approaches and baselines at a constant performance of $0.5$. 
For the 2-qubit Bell state (cf., Fig.~\ref{fig:eval:3}), PPO also manages to evade this local optimum, while SAC shows near-optimal performance. 
However, we look at a different objective of state preparation here, which might be easier, as no exact resemblance of the operations is required. 
Looking at the mean qubits and depth utilized for the GHZ state preparation, SAC, in contrast to the previous Hadamard composition, explores lower operation utilization while exceeding the local optimum for nearly empty circuits (which yields a fidelity of $0.5$). 
This indicates that, while posing a significant challenge, current RL algorithms are capable of exploring hierarchical structures required for building more complex circuit architectures while optimizing for a secondary compactness objective.
Nevertheless, these multi-objective and sparse reward landscapes require explorative capabilities to overcome inherent local minima.

To benchmark more advanced scenarios, Fig.~\ref{fig:eval:2} shows the evaluation results for random state preparation and Toffoli composition.
The ability to quickly put qubits into an arbitrary state is crucial to the success of various approaches, especially in QML using variational quantum circuits. 
Please refer to Fig.~\ref{fig:eval:3} for a benchmark of resembling arbitrary unitary compositions. 
The Toffoli unitary reflects an operation that cannot be directly implemented on most hardware and, therefore, needs to be automatically decomposed. 

For random state preparation, SAC again outperforms all compared baselines with a final mean fidelity of $0.85$.
PPO performs comparably to the GA. 
Notably, all approaches except for TD3 show the capability of learning the objective at hand, even though preparing random states is not considered trivial, which validates our chosen environment design.
Furthermore, the additional compactness objective is shown to be effective, considering that the mean depths predominantly cover only half of the available operations. The better-performing approaches primarily utilize both available qubits. 

Yet, the exploration challenge inherent in the specific actions required to attain more complex targets again is more notable for the Toffoli composition, offering depths up to 63 across three qubits. 
All RL approaches converge to a locally optimal solution with a mean similarity of $0.3$, while both baselines show lower performance around $0.2$. 
Given that an optimal (handcrafted) composition requires a total of 24 concise operations across all three qubits, this unsuccessful behavior is not surprising. 
Looking at both the mean number of qubits and the mean circuit depth reveals primal convergence to empty circuits, indicated by a low depth and low qubit utilization. 
Thus, RL exploits a deficiency in the reward landscape, revealing a local optimum for empty circuits. 
Due to the additional step cost, overcoming this local optimum would first require a performance decrease caused by the use of additional operations. 
On the other hand, the genetic and random baselines produce small yet non-empty circuits that perform even worse regarding their mean similarity to the target unitary.
As expected, random unitary compositing poses a similarly difficult objective, where the best-performing approaches also merely exceed the random baseline.
These results emphasize the need for more directive reward signals to guide policy optimization in such complex, potentially hierarchical tasks with sparse solution landscapes.

Overall, the presented experimental results provide certainty that the proposed \texttt{qcd-gym} environment can be used to learn skills in the domain of QC using RL. 
Yet, current state-of-the-art approaches do not provide sufficient performance for robust and scaleable applicability. 
Besides the proposed objectives and their corresponding metrics, the circuit depth and the number of used qubits have proven to be viable metrics for an in-depth analysis of the learned behavior. 
Across all challenges, SAC has yielded superior results due to its advanced continuous control capabilities.

\section{Conclusion}

We introduced \texttt{qcd-gym}, a framework for benchmarking RL for use in low-level quantum control.  
Furthermore, we motivated and proposed an initial set of objectives for both quantum architecture search and quantum circuit optimization, including the composition of unitary operations and the preparation of quantum states. 
Benchmark results for various state-of-the-art model-free RL algorithms and comparisons to a random and an evolutionary optimization baseline showed RL's competitiveness for QCD while ensuring the soundness of the proposed framework.
Those empirical results also highlighted current challenges to be overcome for RL to be widely applicable for quantum control. 

Those challenges mainly include exploring multi-modal reward landscapes in combination with complex, non-uniform, high-dimensional action spaces and sparse, extinct target requirements. 
While RL research has proposed various approaches to tackle these challenges \citep{hayes2022practical,kim2024guide,li2022hyar,devidze2022exploration}, we believe addressing those quantum-control-specific demands would advance both fields.

Considering the additional objectives, minimizing operation count and circuit depth while maximizing performance separately and applying multi-objective RL concepts might improve the ability to deal with those independent goals. 
Furthermore, smoother reward metrics, including intermediate rewards, are needed to ease exploring the intricate action space for quantum control. 
Those could include more hardware-friendly similarity metrics, also considering specific connectivities or further constraints. 
Also, the proposed state space is currently limited by its scaling exponentially with the number of qubits. 
For an agent to focus on relevant parts of the state space, an attention mechanism or partial observability might be helpful to further decrease the exploration challenge \citep{Altmann23-CROP}. 
In addition, simulations are currently only possible for qubit numbers of $\eta<50$. 
Thus, approximate simulations might be helpful to increase viable circuit sizes.
Future work should also follow up with collecting additional relevant states, unitaries, and Hamiltonians for SP, UC, and HS. 
Also, as previously mentioned, the overall objective might be extended to account for error mitigation.
Furthermore, the action space might be extended to allow for lower-level quantum control, enabling a deeper hardware integration.  

% Future work: extend to pulse shaping?
Overall, however, we believe that \texttt{qcd-gym} provides a profound base for future research on the application of RL to QC to build upon.

\section*{Acknowledgements}
This work is part of the Munich Quantum Valley, which is supported by the Bavarian state government with funds from the Hightech Agenda Bayern Plus.

\balance
\printbibliography
\onecolumn

\section*{Additional Results}

\end{document}